%% file: eprint_dpf2015.tex.new.tex
\documentclass[12pt]{article}
\usepackage{graphicx}
\usepackage{caption}

\usepackage{float}
\usepackage[caption=false]{subfig}

\newcommand\pubnumber{DPF2015-68}
\newcommand\pubdate{\today}

\newcommand{\ETslash}{\ensuremath{E_{\mathrm{T}}\hspace{-1.1em}/\kern0.45em}}

\def\neuniv{Northeastern University \\
Boston, Massachusetts 02115}
\def\Title#1{\begin{center} {\Large #1 } \end{center}}
\def\Author#1{\begin{center}{ \sc #1} \end{center}}
\def\Address#1{\begin{center}{ \it #1} \end{center}}

\newcommand\pubblock{\rightline{\begin{tabular}{l} \pubnumber\\
         \pubdate  \end{tabular}}}
\newenvironment{Abstract}{\begin{quotation}  }{\end{quotation}}
\newenvironment{Presented}{\begin{quotation} \begin{center} 
             PRESENTED AT\end{center}\bigskip 
      \begin{center}\begin{large}}{\end{large}\end{center} \end{quotation}}

\input econfmacros.tex
\begin{document}
\begin{titlepage}
\pubblock

\vfill
\Title{Precision electromagnetic calorimetry at the energy frontier: CMS ECAL at LHC Run 2}
\vfill
\Author{ Andrea Massironi }
\Address{\neuniv}
\vfill
\begin{Abstract}
The CMS electromagnetic calorimeter (ECAL) is a high-resolution, hermetic, 
and homogeneous calorimeter made of 75,848 scintillating lead tungstate crystals.
Following the discovery of the Higgs boson, the CMS ECAL is at the forefront
of precision measurements and the search for new physics in data from the LHC,
which recently began producing collisions at the unprecedented energy of 13 TeV.
The exceptional precision of the CMS ECAL, as well as its timing performance,
are invaluable tools for the discovery of new physics at the LHC Run 2.
The excellent performance of the ECAL relies on precise calibration maintained over time,
despite severe irradiation conditions. A set of inter-calibration procedures
using different physics channels is carried out at regular intervals to normalize
the differences in crystal light transparency and photodetector response between channels,
which can change due to accumulated radiation.
In this talk we present new reconstruction algorithms and calibration strategies which aim to maintain,
and even improve, the excellent performance of the CMS ECAL under the new challenging conditions of Run 2.
\end{Abstract}
\vfill
\begin{Presented}
DPF 2015\\
The Meeting of the American Physical Society\\
Division of Particles and Fields\\
Ann Arbor, Michigan, August 4--8, 2015\\
\end{Presented}
\vfill
\end{titlepage}
\def\thefootnote{\fnsymbol{footnote}}
\setcounter{footnote}{0}

\section{Introduction}
\label{sec:introduction}
The Compact Muon Solenoid (CMS) experiment \cite{Chatrchyan:2008aa} is a general purpose experiment at the Large Hadron Collider (LHC) at CERN,
designed to search for the Standard Model (SM) Higgs boson and for new physics beyond the SM.
Many of these searches involve electrons or photons in the final state,
and the electromagnetic calorimeter (ECAL) plays an essential role in their reconstruction and identification.
The CMS ECAL \cite{CMS:1997ema} has been designed to achieve an excellent energy resolution
which 
is
important for Higgs boson searches with electrons and photons in the final state,
in particular for the H$\to\gamma\gamma$ \cite{Khachatryan:2014ira} and H$\to Z^{(\ast)}Z^{(\ast)}\to 4\ell$ channels \cite{Chatrchyan:2013mxa},
and to guarantee good hermeticity, allowing a precise measurement of the missing transverse energy ($\ETslash$).
The ECAL is a homogeneous and hermetic calorimeter containing 61200 lead tungstate ($\rm PbWO_4$)
scintillating crystals mounted in the barrel (EB), closed at each end by endcaps (EE) each containing 7324 crystals.
The choice of $\rm PbWO_4$ with a radiation length $X_0=0.89$ cm and a Moliere radius $R_0$=2.19 cm ensures the
compactness of the detector and the radiation hardness necessary to cope with the harsh environment of the LHC.
The scintillation light is detected by avalanche photodiodes (APDs) in EB and by vacuum phototriodes (VPTs) in EE.

\section{Energy reconstruction}
\label{sec:energyreco}
The electrical signal from the photodetectors is amplified and shaped by a multi-gain preamplifier.
The output is digitized by a 12 bit ADC running at 40 MHz,
which records ten consecutive samples
used to reconstruct the signal amplitude.

\subsection{Signal pulse reconstruction}
\label{sec:amplitudereco}
During LHC Run I a digital filtering algorithm was used, where the signal amplitude is estimated as the linear combination of the $N=10$ samples, $S_i$:
\begin{equation}
\hat{\cal A}=\sum_{i=1}^{N} w_i \times S_i
\end{equation}
where the weights $w_i$ are calculated by minimizing the variance of $\hat{\cal A}$ \cite{Bruneliere:2006ra},
thus providing an optimal filtering of the electronics noise,
that is estimated on an event-by-event basis by averaging the first three digitized pedestal-only samples.

\begin{figure}[!t]
  \begin{center}
    \subfloat[]{\includegraphics[width=0.4\textwidth]{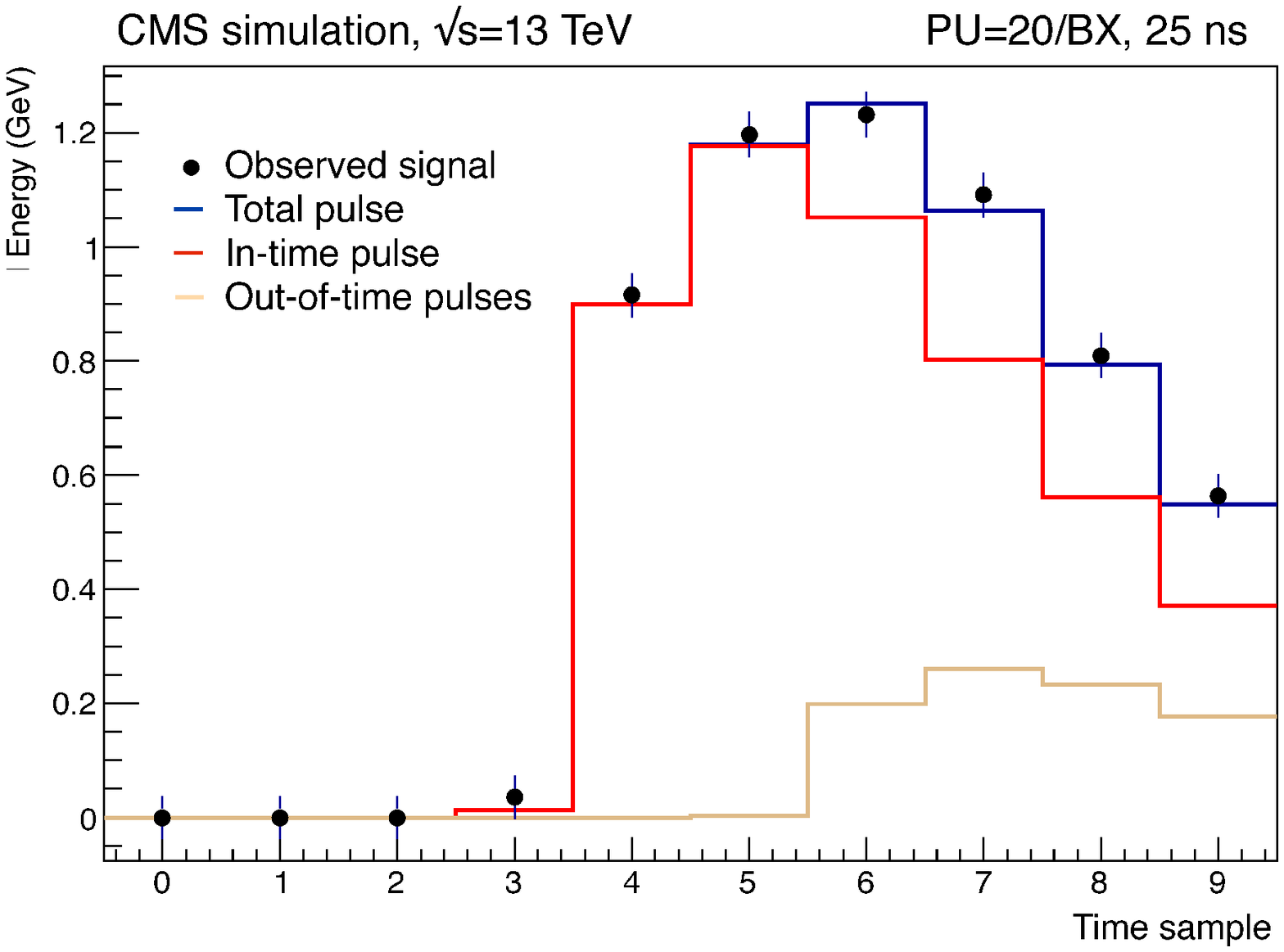} \label{fig:multifit_EB}}
    \subfloat[]{\includegraphics[width=0.4\textwidth]{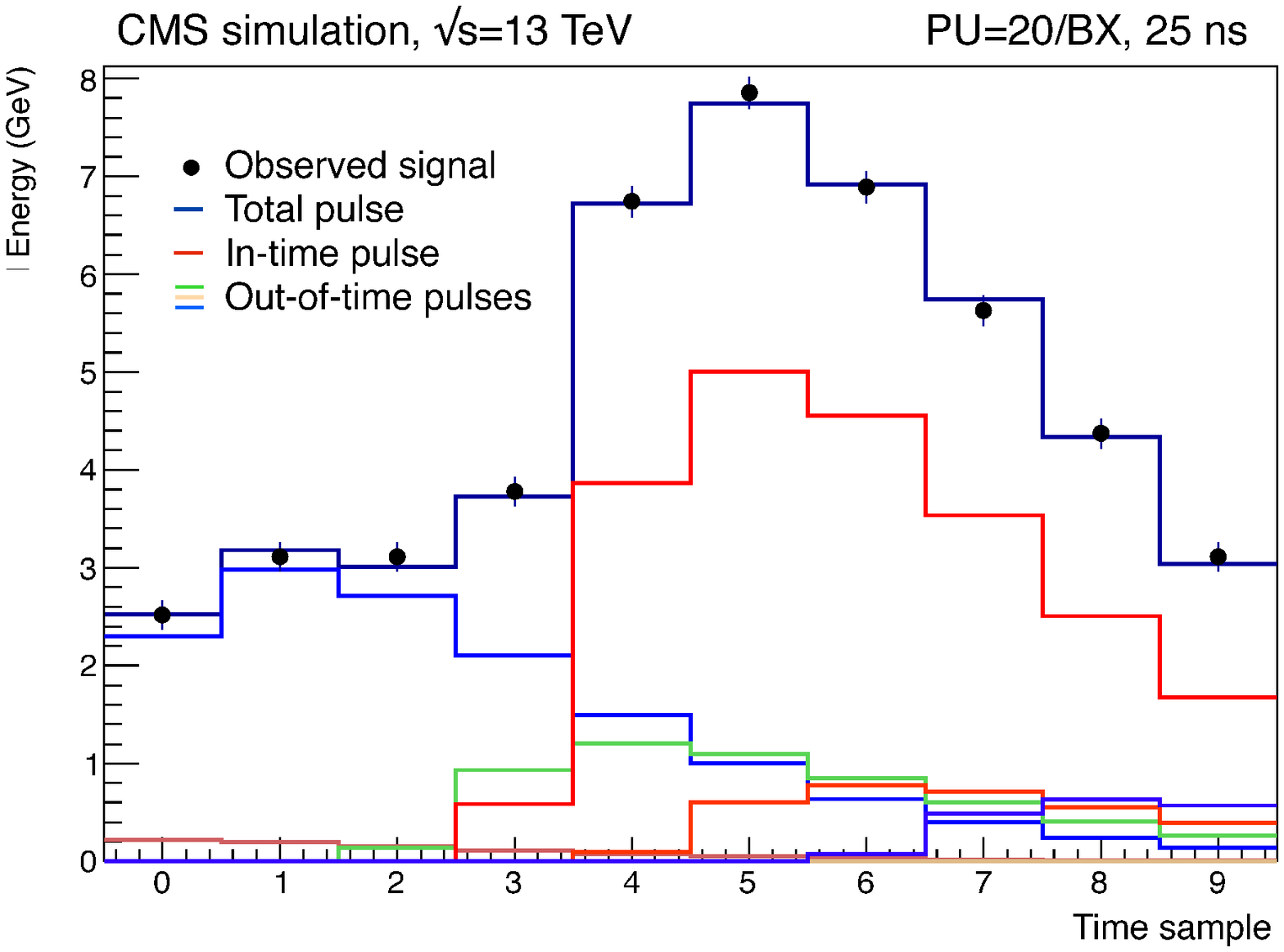} \label{fig:multifit_EE}}
    \caption{
       Two examples of fitted pulses for simulated events with 20 average pileup interactions and 25 ns bunch spacing,
       for a signal in the barrel (a) and in the endcaps (b).
       Dots represent the 10 digitized samples, the red distributions (other light colors)
       represent the fitted in-time (out-of time) pulses with positive amplitudes.
       The dark blue histograms represent the sum of all the fitted contributions.  \label{fig:multifits}
       }
  \end{center}
\end{figure}

The conditions that will be experienced during LHC Run II place particular requirements on the ECAL pulse reconstruction algorithms.
The instantaneous luminosity is expected to increase by a factor of two compared to Run I,
thus the number of inelastic collisions per LHC bunch crossing (pileup) will also increase,
with an average of $\sim$40 collisions per bunch crossing expected during the highest intensity LHC collisions in 2015.
In addition the spacing between colliding bunches is reduced from 50ns to 25ns in 2015,
increasing the level of out-of-time pileup.
Several methods have been investigated to mitigate the effect of pileup, maintaining optimal noise filtering.
A template fit with multiple components, termed ``multi-fit'', is now being used in Run II. 
The multi-fit algorithm estimates the in-time signal amplitude and up to 9 out of time amplitudes by minimization of the $\chi^2$, given by
\begin{equation}
\label{eqn:chi2multifit}
\chi^2 = \sum_{i=1}^{N} \frac{\left(\sum_{j=1}^M {\cal A}_{j}p_{ij} - S_i \right)^2}{\sigma^2_{S_i}}
\end{equation}
where ${\cal A}_{j}$ are the amplitudes of up to $M=10$ interactions.
The pulse templates $\mathbf{\vec p}_j$ for each bunch crossing $j$ have the same shape,
but are shifted in time by multiples of 25 ns within a range of -5 to +4 bunch crossings (BX) around the time of the in-time signal (BX=0).
The pulse templates for each crystal are measured from low pileup $pp$ collision data recorded by CMS at the beginning of 2015. 
The total electronic noise and its associated covariance matrix, $\sigma_{S_i}$ are measured from dedicated pedestal runs.
The technique of Non-Negative-Least-Squares \cite{nnls} is used to perform the $\chi^2$ minimization of Eq.~\ref{eqn:chi2multifit}
with the constraint that the fitted amplitudes are all positive.
Examples of one fit for signals in the barrel and in the endcaps are shown in Fig.~\ref{fig:multifits},
for an average pileup of 20 and for 25 ns bunch spacing.

The predicted improvement in energy resolution with respect the Run I reconstruction algorithm for collisions with 25 ns bunch spacing
is substantial especially for low $p_T$ photons and electrons,
given the larger contribution of pileup to the total energy estimate, 
and is still significant for those at high $p_T$ ($p_T>$50 GeV). 

\subsection{Clustering algorithms}
\label{sec:clustering}
Electrons and photons deposit their energy over several ECAL channels
and the presence of material in front of ECAL causes conversions of photons
and bremsstrahlung from electrons, so that the radiated energy
is spread along $\phi$ by the magnetic field.
Clustering algorithms are used to collect the energy deposits in ECAL,
including the contributions from this radiated energy.
The electron or photon energy is then estimated as:
\begin{equation}
\label{eqn:energy}
E_{e,\gamma} = F_{e,\gamma}\left[G \times \sum_{i=crystal}\left( C_i \times S_i(t) \times {\cal A}_i \right) + E_{ES}\right]
\end{equation}
where the sum is performed over all the clustered crystals.
The amplitude measured in the $i$-th crystal is labeled by ${\cal A}_i$,
while $S_i(t)$ is a time dependent correction that accounts for time variation of the channel response
due to changes in crystal transparency.
The $C_i$ parameter is a relative calibration constant that takes into account differences
among crystals for light yields and photodetector response
and $G$ is a scale factor converting the digital scale into GeV.
For clusters in the endcap region the corresponding energy in the preshower ($E_{ES}$) is added.
Finally $F_{e,\gamma}$ is a particle dependent correction applied to the clustered energy
that accounts for biases in the energy reconstruction related to the geometry of the detector,
the upstream material, 
the electromagnetic shower leakage
and the clustering of energy emitted by bremsstrahlung or photon conversions.
Residual data/simulation differences are accounted for by in-situ measurements
using several physics processes producing electrons and photons in the final state: $\pi^0,\eta\to\gamma\gamma$, $W\to e\nu$, $Z\to e^+e^-$.

\section{Energy calibrations}
\label{sec:energycalibration}

The calibration procedure of the ECAL response in Run II uses the techniques developed during Run I~\cite{Chatrchyan:2013dga},
while reoptimizations of the calibration streams have been necessary
to cope with the increased rates and larger data volumes that are expected.

\subsection{Corrections for time-dependent response changes}
\label{sec:timedependence}
Ionizing radiation creates color centers in the crystals reducing their transparency
and therefore reducing their measured response to the deposited energy.
The color centers partially anneal with thermal energy such that the loss in transparency depends on the dose rate,
which varies with $\eta$, and a partial recovery of transparency is observed in the absence of radiation.
Changes in crystal transparency and photodetector response are measured
and corrected using a dedicated laser monitoring system \cite{Anfreville:2007zz} 
which injects laser light
into each crystal.
The measured changes in response measured by the laser system ($R/R_0$) are related to
changes in the scintillation signal ($S/S_0$) by means of a power law:
$\frac{S}{S_0} = \left(\frac{R}{R_0}\right)^\alpha$, 
where $\alpha ~\sim 1.5$.
By the end of LHC Run I in 2012 the loss in response was up to 6\% in EB and up to 30\% in the EE region
up to $\vert\eta\vert<2.5$
and reaching 70\% in the most forward EE regions.

The corrections for response loss 
were validated with collisions data,
by examining the stability of the reconstructed invariant mass of $\pi^0$ decays
and using the ratio of $E/p$ for isolated electrons from $W\to e\nu$ and $Z\to e^+e^-$ decays, 
where $E$ is the energy measured in the calorimeter and $p$ is the momentum measured in the tracker.

\begin{figure}[!t]
  \begin{center}
    \subfloat[]{\includegraphics[width=2.3in]{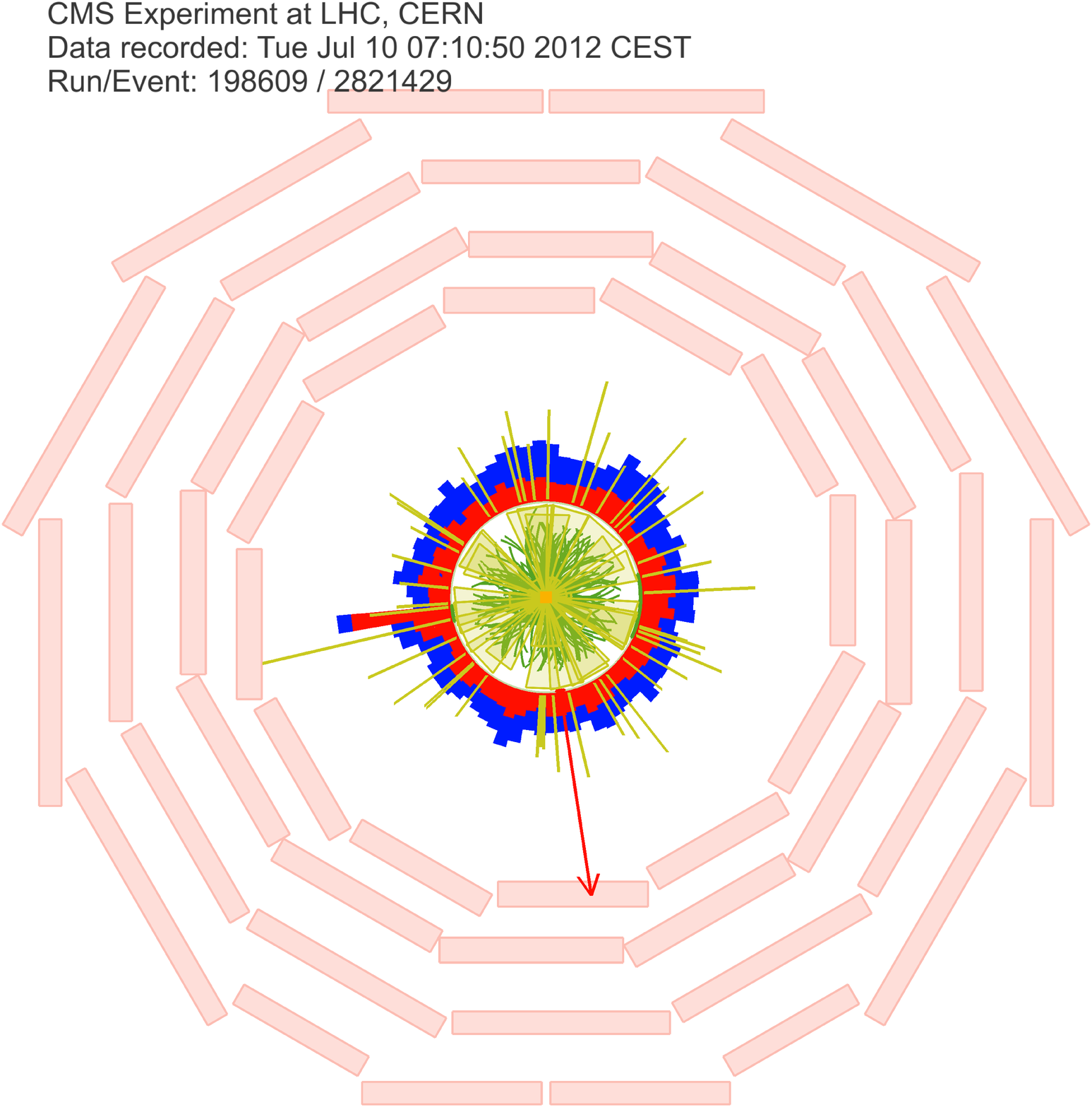}\label{fig:elestream_before}}
    \subfloat[]{\includegraphics[width=2.3in]{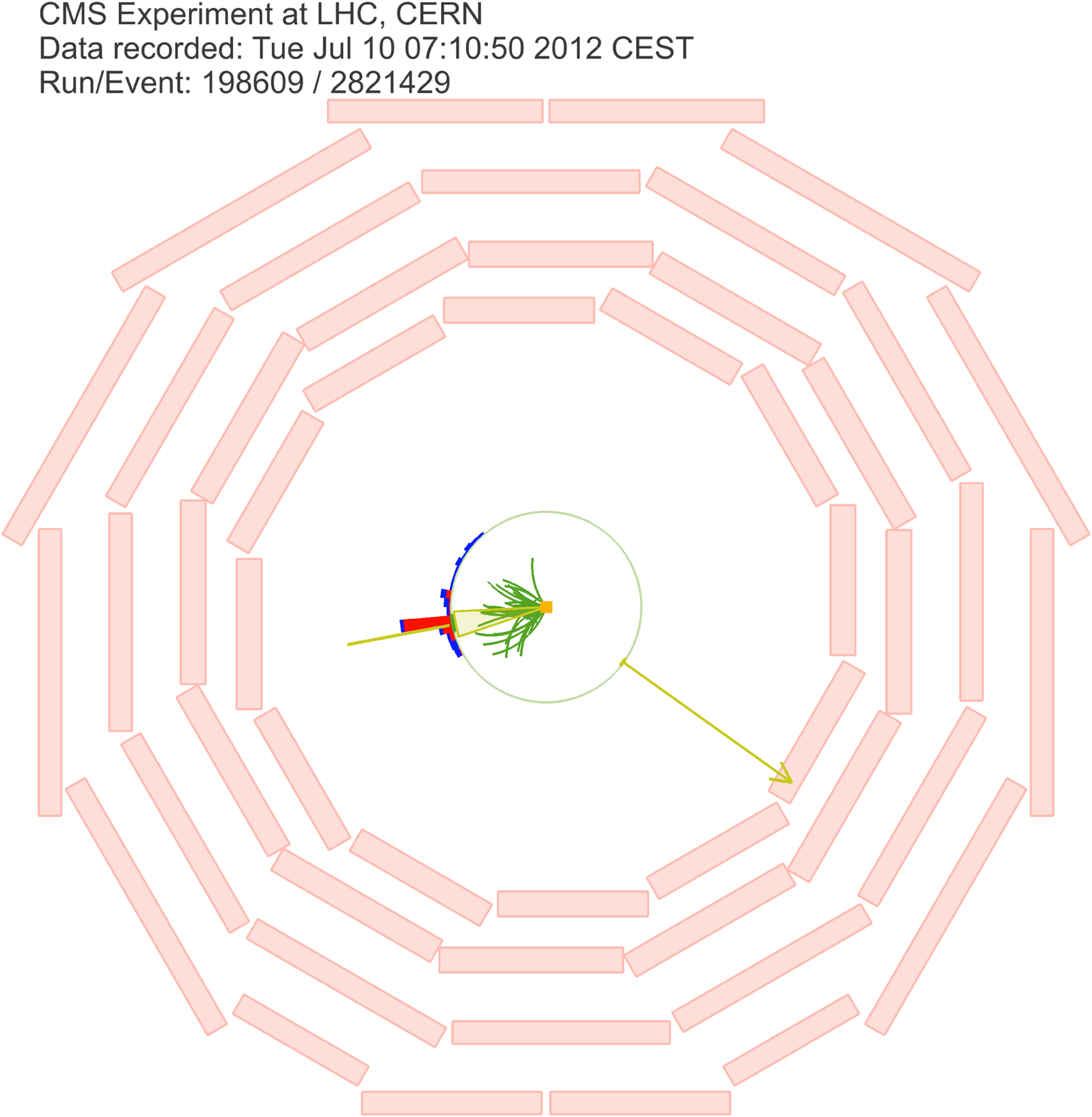}\label{fig:elestream_after}}
    \caption{
      Event displays of a $W\to e\nu$ event reconstructed offline,
      with the full event content (a) and after the event content has been reduced
      to save only the hits around the electron (a).
      On Fig. (a), the red arrow shows the reconstructed $\ETslash$, while
      on Fig. (b), the yellow arrow shows the $\ETslash$ as estimated
      using information available online at trigger level.
      \label{fig:elestream}
      }
  \end{center}
\end{figure}

For Run II one of the main challenges is to maximise the number of $W\to e\nu$ decays available
for calibration purposes give the limited bandwidth available to single electron triggers in 2015.
A dedicated data stream has been developed to reduce the output bandwith per event
by recording only the hits from the tracker and from the ECAL regions that are traversed
by the reconstructed electron.
The event size is typically reduced by a factor of 5 to 10.
Event displays of a $W\to e \nu$ interaction 
reconstructed with the full event content,
and with the reduced event content are shown
in Fig.~\ref{fig:elestream_before} and \ref{fig:elestream_after} respectively.

\subsection{Intercalibrations and energy scale}
\label{sec:intercalibrations}

The relative calibration of the crystals (parameter $C_i$ in Eq.~\ref{eqn:energy})
is obtained from LHC collisions data using several independent methods,
and the resulting constants are combined to provide one number per crystal.
These methods include the use of azimuthal symmetry of the energy flow in minimum bias events (``$\phi$ symmetry''),
the invariant mass of photon pairs from $\pi^0$ and $\eta$ decays,
and the $E/p$ ratio of isolated electrons from $W\to e\nu$ and $Z\to e^+e^-$ decays.
All these methods
are being used to provide the first results with the data collected at 13 TeV.

The combined relative calibration for Run I 
was obtained from the mean of the individual corrections
weighted by their respective precisions. 
In the region $\vert\eta\vert>2.5$, beyond the tracker acceptance,
the $E/p$ method can not be used and the high pileup prevents the reconstruction of the
invariant mass peak from $\pi^0$ or $\eta$, therefore only $\phi$-symmetry
and $Z\to e^+e^-$, with one electron and one electromagnetic cluster,
information is available.
The precision of the combination is $\sim$0.3\% ($\sim$0.7\%) in the central (outer) barrel,
and $\sim$1.5\% in the endcaps.

\section{ECAL alignment}

Electron identification relies upon matching the measurements in the ECAL and the Tracker
to better than 0.02 radians in $\phi$ and 4$\times$10$^{-3}$ units in $\eta$~\cite{P06005}.
In addition, the accurate position measurement of photons impacting on the calorimeter
is used to determine their direction with respect to the collision vertex~\cite{P08010}.
The accuracy of the measurement of the opening angle between the two decay photons
from the Higgs boson contributes to its reconstructed invariant mass resolution.
The precise alignment of ECAL within CMS is therefore necessary 
to achieve the required position measurement resolution.
The relative alignment of the ECAL with respect to the tracker
is performed using electrons from W and Z boson decays.
During Run I the required position resolution has been achieved~\cite{P06005}.
A new evaluation of the ECAL alignment is being performed with the 13 TeV data.

\section{Conclusion}
\label{sec:conclusions}
The electromagnetic calorimeter of CMS has demonstrated excellent performance during LHC Run I,
and it has been essential in the discovery of the Higgs boson and in the determination of its properties.
The harsh radiation environment of the LHC has required a continuous effort in the operation,
monitoring and calibration of the calorimeter. 

For the Run II at $\sqrt{s}=13$ TeV, with an increased instantaneous luminosity,
an almost doubled number of pileup interactions and the reduced bunch spacing from 50 ns to 25 ns,
the environment is even more challenging.
A new ECAL amplitude reconstruction have been developed to cope with 
the increased pileup.
The calibration procedures and techniques are
build upon the techniques developed during Run II
and a re-optimization of the streams has been performed
during the shutdown period between Run I and Run II.
These improvements will allow ECAL to maintain excellent performance during LHC Run II,
both for precision physics including Higgs boson properties and other SM measurements,
and for the searches of BSM particles with decay chains involving electrons and photons.

\end{document}

%% file: econfmacros.tex
\def\beq{\begin{equation}}
\def\eeq#1{\label{#1}\end{equation}}
\def\eeqn{\end{equation}}

\def\beqa{\begin{eqnarray}}
\def\eeqa#1{\label{#1}\end{eqnarray}}
\def\eeqan{\end{eqnarray}}

\let\bar=\overbar

\def\Dslash{\not{\hbox{\kern-4pt $D$}}}
\def\dslash{\not{\hbox{\kern-2pt $\del$}}}

\def\msb{{\bar{\ssstyle M \kern -1pt S}}}

